\documentclass[twoside,slac_one]{revtex4}
\usepackage{graphicx}
\usepackage{fancyhdr}
\usepackage{amsmath} 
\usepackage{bm}
\usepackage{amsxtra}
\usepackage{amssymb}
\usepackage{amsthm}
\usepackage{latexsym}
\usepackage{lscape}

\begin{document}

\title{Efficiency measurement of b-tagging algorithms developed by the CMS experiment}

%

\author{Saptaparna Bhattacharya for the CMS collaboration}
\affiliation{Department of Physics and Astronomy, Brown University, Providence, RI, USA}

\begin{abstract}
Identification of jets originating from b quarks (b-tagging) is a key element of many physics analyses at the LHC. Various algorithms for b-tagging have been developed by the CMS experiment to identify b-tagged jets with a typical efficiency between 40$\%$ and 70$\%$ while keeping the rate of misidentified light quark jets between 0.1$\%$ and 10$\%$. An important step, in order to be able to use these tools in physics analysis, is the determination of the efficiency for tagging b-jets. Several methods to measure the efficiencies of the life-time based b-tagging algorithms are presented. Events that have jets with muons are used to enrich a jet sample in heavy flavor content. The efficiency measurement relies on the transverse momentum of the muon relative to the jet axis or on solving a system of equations which incorporate two uncorrelated taggers. Another approach uses the number of b-tagged jets in top pair events to estimate the efficiency. The results obtained in 2010 data and the uncertainties obtained with the different techniques are reported. The rate of misidentified light quarks have been measured using the ``negative'' tagging technique.
\end{abstract}

\maketitle

\thispagestyle{fancy}


\section{Introduction}

B tagging or the identification of b-jets is of crucial importance in event topologies involving b-quarks. Many standard model processes entail b-quark production in the intermediate state, for example, in top physics b-tagging is imperative to distinguish between signal and background processes. Higgs physics is heavily b-tagging dependent when the Higgs primarily decays to $b {\bar{b}}$ pairs at a mass of 120 GeV. Hence, for such processes, the efficiency of tagging b-jets is an important variable in the analysis. The CMS detector has performed remarkably well. There is good agreement between data and simulations. However, b-tagging is a complex tool that relies on many aspects of detector performance and hence it is essential to measure the b-tagging efficiency on data and not rely exclusively on input from simulations.

The algorithms for b-jet identification utilize several salient features of B hadron decays. B hadrons have a relatively high lifetime of $\sim$1.5 ps (c$\tau$ = 450 $\mu$m). They have a mass of $\sim$ 5.2 GeV, which is higher than the mass of the light quarks. They typically tend to decay into a large number of charged particles, the average decay multiplicity being $\sim 5$. Due to the high mass, the fragmentation is hard, hence the $p_{T}$  of decay products is high. The semi-leptonic branching ratio of B hadrons is $\sim$ 11$\%$ for each lepton flavor. This branching ratio is as high as $\sim$ 20$\%$ when $b \rightarrow c$ cascade decays are taken into account. These properties allow b-jets to be distinguished from light jets (u, d, s) or gluon jets and to a lesser extent c-jets.
  
\section{B tagging Algorithms}

The inputs to b-tagging are particle flow jets  \cite{Particle_Flow_paper}, charged particle tracks and vertices, both primary and secondary. The jets are reconstructed by the anti-$k_{T}$ clustering method, with a cone radius parameter of $\Delta R$=0.5, where $\Delta R$ is defined in terms of the azimuthal angle $\phi$ and pseudorapidity $\eta$ as $\Delta R = \sqrt{{\Delta \eta}^2 + {\Delta \phi}^2}$. The tracks are reconstructed with a Kalman Filter based method \cite{Kalman}. The vertices are reconstructed from tracks compatible with the beam spot using the Adaptive Vertex Fitter algorithm \cite{AVF}. The output of the b-tagging algorithms is a discriminator. This is a variable which is sensitive to the flavor content of the jet and is computed from tracks associated with the jets. The next step is to choose a working point. A loose operating point implies a 10$\%$ light quark fraction, while medium and tight correspond to 1$\%$ and 0.1$\%$ light quark fractions respectively.

\begin{figure}[htp] 
\centering
\includegraphics[width=3.00in, height=2.2in]{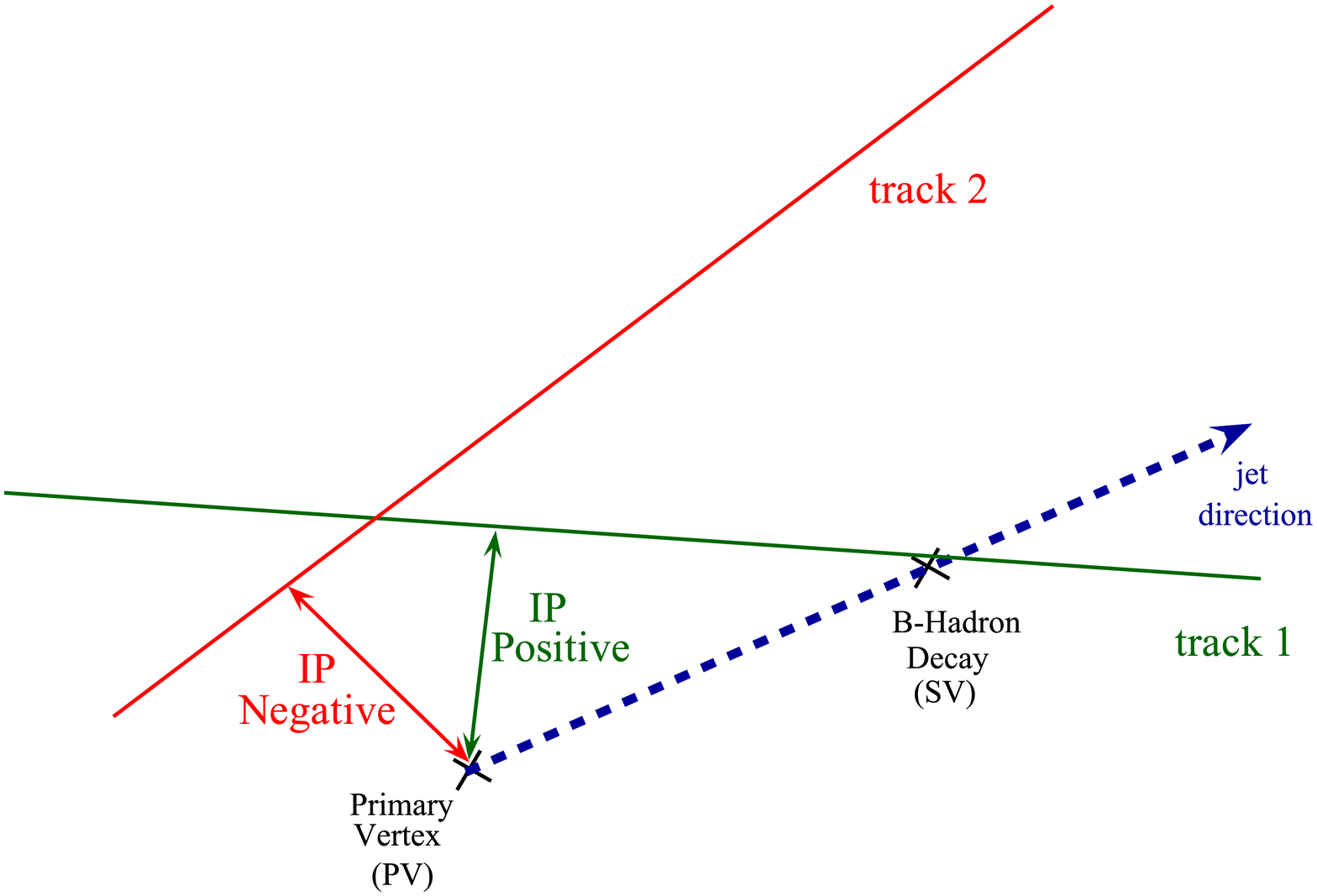}
\caption{Definition of positive and negative impact parameters}
\label{impact_parameter}
\end{figure}

The algorithms for b-jet identification utilize the unique features of B hadron decays. The impact parameter (IP) is defined as the two dimensional or three dimensional distance between the track and the vertex at the point of closest approach as shown in Fig.~ \ref{impact_parameter}. Since the uncertainty, $\sigma_{IP}$, varies with the number of tracks, the preferred b-tagging variable is $IP/\sigma_{IP}$ . The lifetime based taggers rely on tracks with large impact parameters or on the presence of a reconstructed secondary vertex within a jet. Track Counting (TC) and Jet Probability (JP) are impact parameter based taggers. The TC discriminator is based on finding $N$ tracks with $IP/\sigma_{IP} > S$, where $S$ is a threshold. In the high efficiency (HE) version of this tagger, the value of $N$ is set at two, while the high purity (HP) tagger utilizes the first three tracks. The HP version of the tagger, hence, has a lower b-tagging efficiency due to the application of a stringent cut. Consequently, the mis-tag rate is also low. The JP tagger combines information from all tracks and computes the probability of these tracks to come from the primary vertex.  An alternate version of the JP tagger used in analyses is based on enhancing the b flavor content by associating a higher weight to the four most displaced tracks. This form of the JP tagger is analogous to a HP version of the tagger. 
The next set of b tagging algorithms involve a secondary vertex in ${\bf B}$ hadron decays. The simple secondary vertex (SSV) tagger is based on the reconstruction of at least one secondary vertex. The discriminating variable for this tagger is obtained from the significance of the 3D flight distance. SSVHE is obtained by associating two tracks with the vertex, while SSVHP relies on three tracks associated with the vertex. 

These taggers are simple taggers that do not require calibration, therefore, ideal for early data taking.  In addition to these taggers, the complex secondary vertex tagger (CSV) is used. This tagger uses various track and vertex information combined through a multi-variate technique.\\

\subsection{Efficiency measurement from muon-jet events : $p_{Trel}$ method}

\begin{figure}[htp]
\centering
 \includegraphics[width=1.4in, height=1.8in]{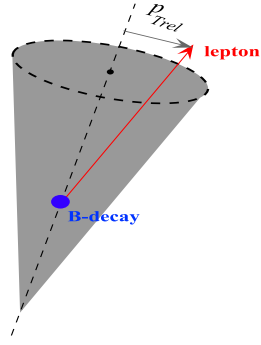} 
 \caption{$p_{Trel}$ is defined as the transverse momentum of the muon with respect to the jet direction.}
  \label{ptrel}
 \end{figure}

The $p_{Trel}$ method utilizes semi-leptonic B hadron decays giving rise to b-jets that contain a muon (``muon jet''). $p_{Trel}$ is defined as the transverse momentum of the muon with respect to the jet direction as pictorially described in Fig.~\ref{ptrel}. Due to the high b quark mass, $p_{Trel}$ is larger for muons from B hadron decays. A sample, with an enhanced b-jet purity, is constructed by asking for two reconstructed jets : the muon-jet and another fulfilling the b-tagging criterion. The $p_{Trel}$ spectra for muon jets originating from $b$, $c$ and light flavor partons are obtained from simulations. $f_{b}^{tag}$ ($f_{b}^{untag}$) are defined as fractions of jets that pass (fail) the b-tagging requirement. From the $p_{Trel}$ spectra of $b$ and non-$b$ ($c$ + light flavor jets), these fractions are extracted with a maximum likelihood fit. The fractions and the total number of tagged and untagged muon jets ($N_{data}^{tag}$, $N_{data}^{untag}$) are used to calculate the efficiency: $\varepsilon_{b}^{tag} = \frac{f_{b}^{tag}.N_{data}^{tag}}{f_{b}^{tag}.N_{data}^{tag} + f_{b}^{untag}.N_{data}^{untag}}$. The plots of the fits to the $p_{Trel}$ distributions are in Fig.~\ref{ptrel_fits}.

\begin{figure}[htp] \label{ptrel_fits}
\centering
\includegraphics[width=3.0in, height=2.2in]{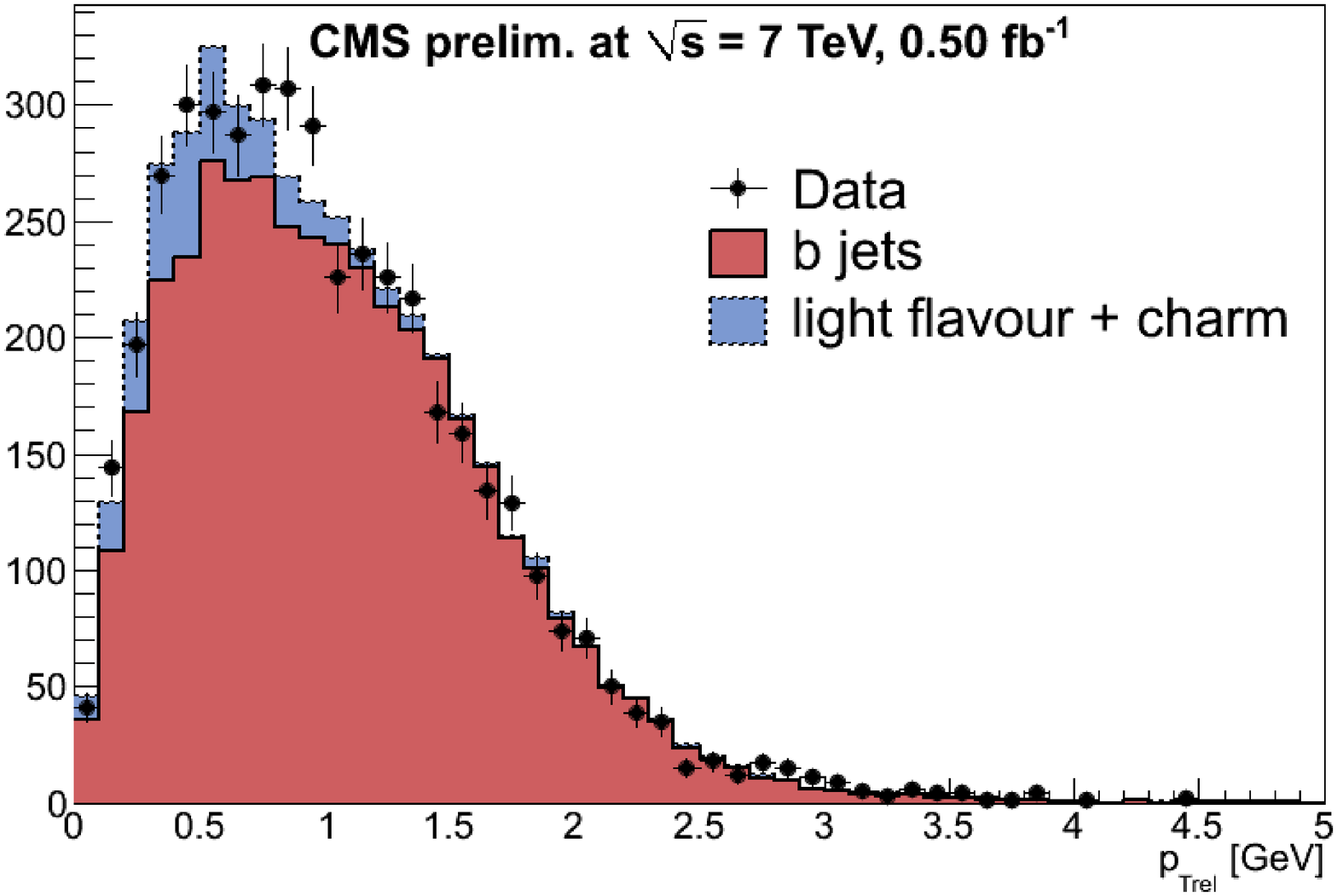}
\includegraphics[width=3.0in, height=2.2in]{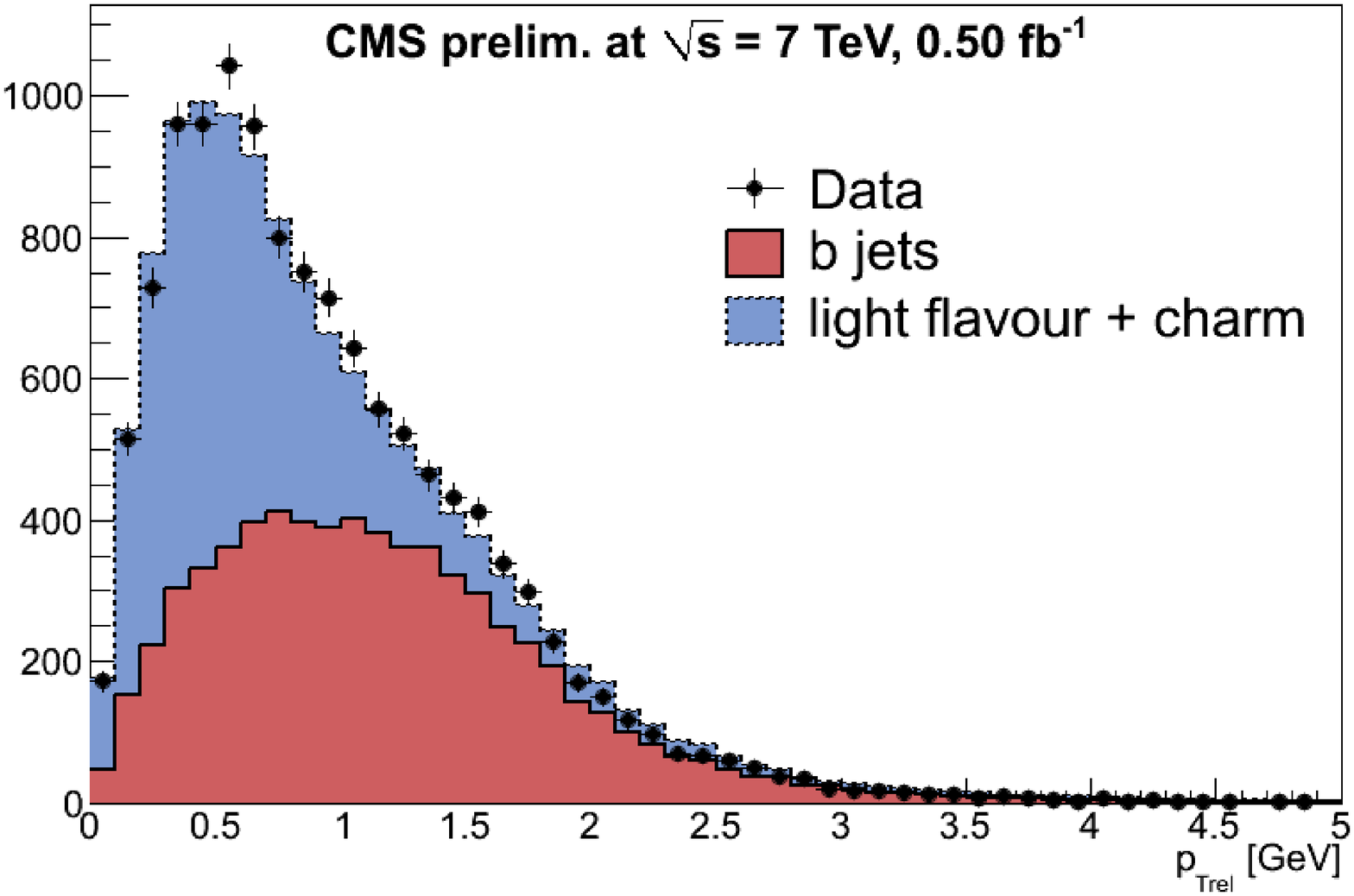}  
\caption {Fits of the muon $p_{Trel}$ distributions to b and light flavor templates for jets containing muons that (left) pass or (right) fail the b-tagging algorithm: SSVHPT (Simple Secondary Vertex High Purity Tight Operating Point). The fractions and the total yields ($N_{data}^{tag}$, $N_{data}^{untag}$) are used to calculate the efficiency.}
 \label{ptrel_fits}
\end{figure}

\subsection{``System 8"}

``System 8'' is a data driven method with minimal dependence on simulations. System8, like the $p_{Trel}$ method, takes advantage of semi-leptonic B hadron decays. It is applied to a sample of muon jet events. A system of 8 non-linear equations are set up and solved using numerical methods. Two data samples are used: 

\begin{itemize}
\item The muon jet+ away-jet sample : Contains two reconstructed jets and a muon within $\Delta R < 0.4$ of one of the jets. The highest $p_{T}$ muon is taken when there exist more muons in the jet. If there exist two jets with muons in them in an event, both are counted as muon jets.

\item The muon jet+tagged-away-jet sample :  This sample is created by tagging a b quark in the away jet. Since b quarks are produced in pairs a b quark can be tagged in the same event in another jet.

\end{itemize} 
 
The first two equations, hence are:

\begin{eqnarray} 
n = n_b + n_{cl} \\
p = p_b + p_{cl} 
\end{eqnarray}

Here, $(n,p)$ are the muon-in-jets in each sample.\\
 
Two different taggers are used:  A test tagger (``tag") which in this case is chosen to be a lifetime based tagger and a cut on $p_{Trel}$. This choice is dictated by the requirement that these taggers be minimally correlated.\\

Hence the next set of equations are:
\begin{eqnarray} 
n^{tag} = \varepsilon_{b}^{tag}n_{b} + \varepsilon_{cl}^{tag}n_{cl} \\
~~~p^{tag} = \beta_{12} \varepsilon_{b}^{tag}p_{b} + \alpha_{12} \varepsilon_{cl}^{tag}p_{cl} 
\end{eqnarray}

Here, $(n^{tag}, p^{tag})$ are lifetime tagged.

\begin{eqnarray} 
n^{p_{Trel}} = \varepsilon_{b}^{p_{Trel}}n_{b} + \varepsilon_{cl}^{p_{Trel}}n_{cl} \\
~~~p^{p_{Trel}} = \beta_{23} \varepsilon_{b}^{p_{Trel}}p_{b} + \alpha_{23}\varepsilon_{cl}^{p_{Trel}}p_{cl} 
\end{eqnarray}

Here, $(n^{p_{{Trel}}}, p^{p_{Trel}})$ are obtained by applying a cut on the $p_{Trel}$ distribution.

\begin{eqnarray}
n^{tag,p_{Trel}}  =  \beta_{13}\varepsilon_{b}^{tag}\varepsilon_{b}^{p_{Trel}}n_{b} + \alpha_{13}\varepsilon_{cl}^{tag}\varepsilon_{cl}^{p_{Trel}}n_{cl}\\
~~~p^{tag,p_{Trel}}  =  \beta_{123} \varepsilon_{b}^{tag}\varepsilon_{b}^{p_{Trel}}p_{b} + \alpha_{123} \varepsilon_{cl}^{tag}\varepsilon_{cl}^{p_{Trel}}p_{cl}
\end{eqnarray}
 
The last set of equations are a result of the application of both tags.\\
 
The correlation factors are $(\alpha_{12}, \beta_{12}, \alpha_{23}, \beta_{23}, \alpha_{13}, \beta_{13}, \alpha_{123}, \beta_{123})$  obtained from simulations. They are defined as:\\

\begin{equation} \beta_{12} = \frac{{\varepsilon}_{b}^{tag} \text{from muon jet+tagged-away-jet sample}}{{\varepsilon}_{b}^{tag}\text{from muon-jet+away-jet sample}}\end{equation}
\begin{equation} \alpha_{12} = \frac{{\varepsilon}_{cl}^{tag} \text{from muon jet+tagged-away-jet sample}}{{\varepsilon}_{cl}^{tag}\text{from muon-jet+away-jet sample}}\end{equation}
 \begin{equation} \beta_{23} = \frac{{\varepsilon}_{b}^{p_{Trel}} \text{from muon jet+tagged-away-jet sample}}{{\varepsilon}_{b}^{p_{Trel}}\text{from muon-jet+away-jet sample}}\end{equation}
 \begin{equation} \alpha_{23} = \frac{{\varepsilon}_{cl}^{p_{Trel}} \text{from muon jet+tagged-away-jet sample}}{{\varepsilon}_{cl}^{p_{Trel}}\text{from muon-jet+away-jet sample}}\end{equation}

\begin{equation} \beta_{13} = \frac{{\varepsilon}_{b}^{tag, p_{Trel}}}{{\varepsilon}_{b}^{tag}{\varepsilon}_{b}^{p_{Trel}}}
\hspace{0.1in} \text{and} \hspace{0.1in}
\alpha_{13} = \frac{{\varepsilon}_{cl}^{tag, p_{Trel}}}{{\varepsilon}_{cl}^{tag}{\varepsilon}_{cl}^{p_{Trel}}}\end{equation}

for the muon jet and away-jet sample and,

\begin{equation} \beta_{123} = \frac{{\varepsilon}_{b}^{tag, p_{Trel}}}{{\varepsilon}_{b}^{tag}{\varepsilon}_{b}^{p_{Trel}}}
\hspace{0.1in} \text{and} \hspace{0.1in}
\alpha_{123} = \frac{{\varepsilon}_{cl}^{tag, p_{Trel}}}{{\varepsilon}_{cl}^{tag}{\varepsilon}_{cl}^{p_{Trel}}}\end{equation}

for the muon jet and tagged-away-jet sample.\\

These definitions are obtained by writing the left hand side of the equations in terms of a composite efficiency term (${\varepsilon}_{b}^{tag, p_{Trel}}$) and equating the $b$ and $c$ and light jet terms on each side of the equation. These correlation factors are the only variables that are obtained from simulations, hence, justifying the claim that this method is data-driven.

\subsection{Measured $b$-tagging efficiencies}

\begin{figure}[htp]
\centering
\includegraphics[height=3.0in, width=2.8in]{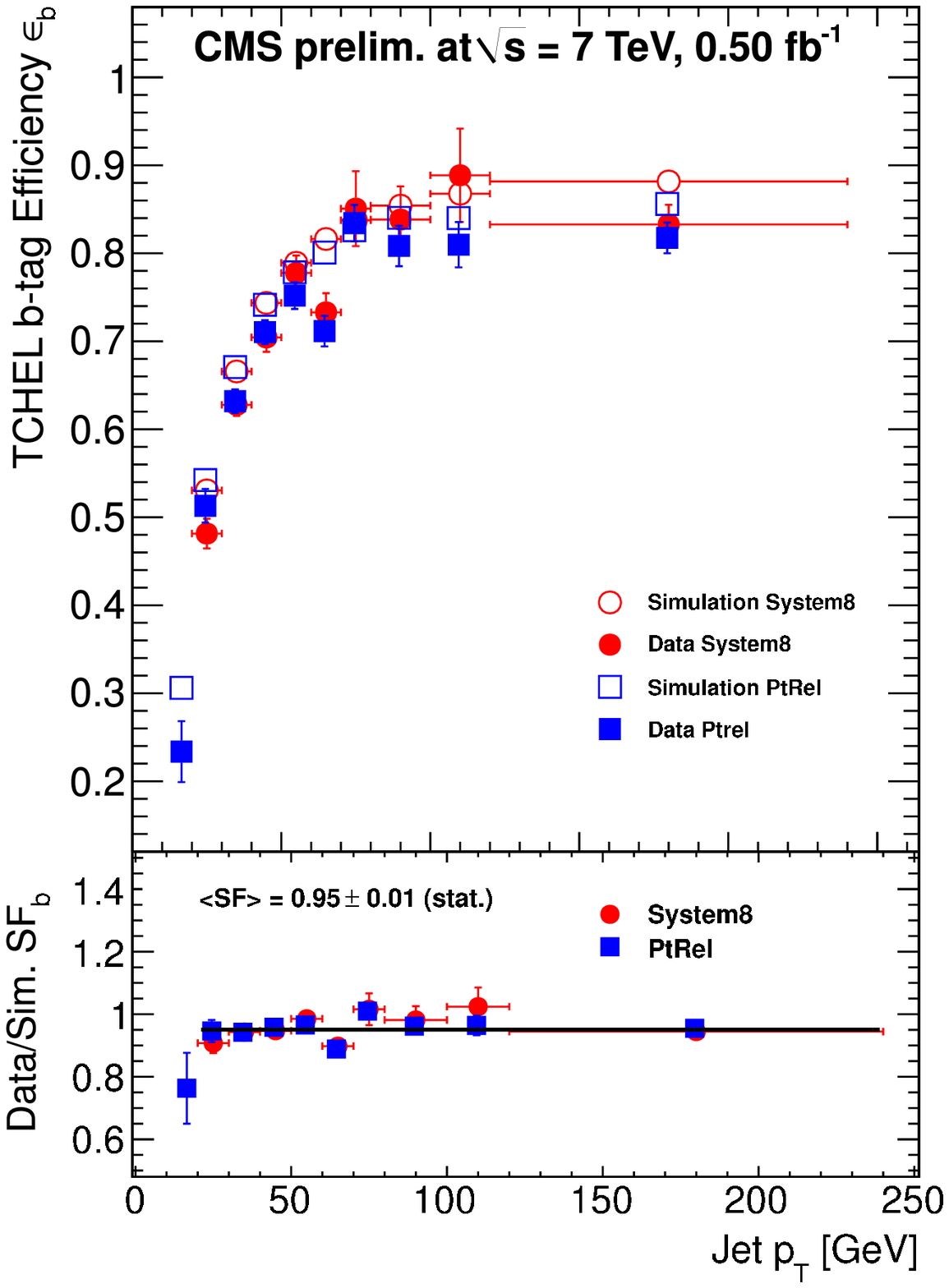}
\includegraphics[height=3.0in, width=2.8in]{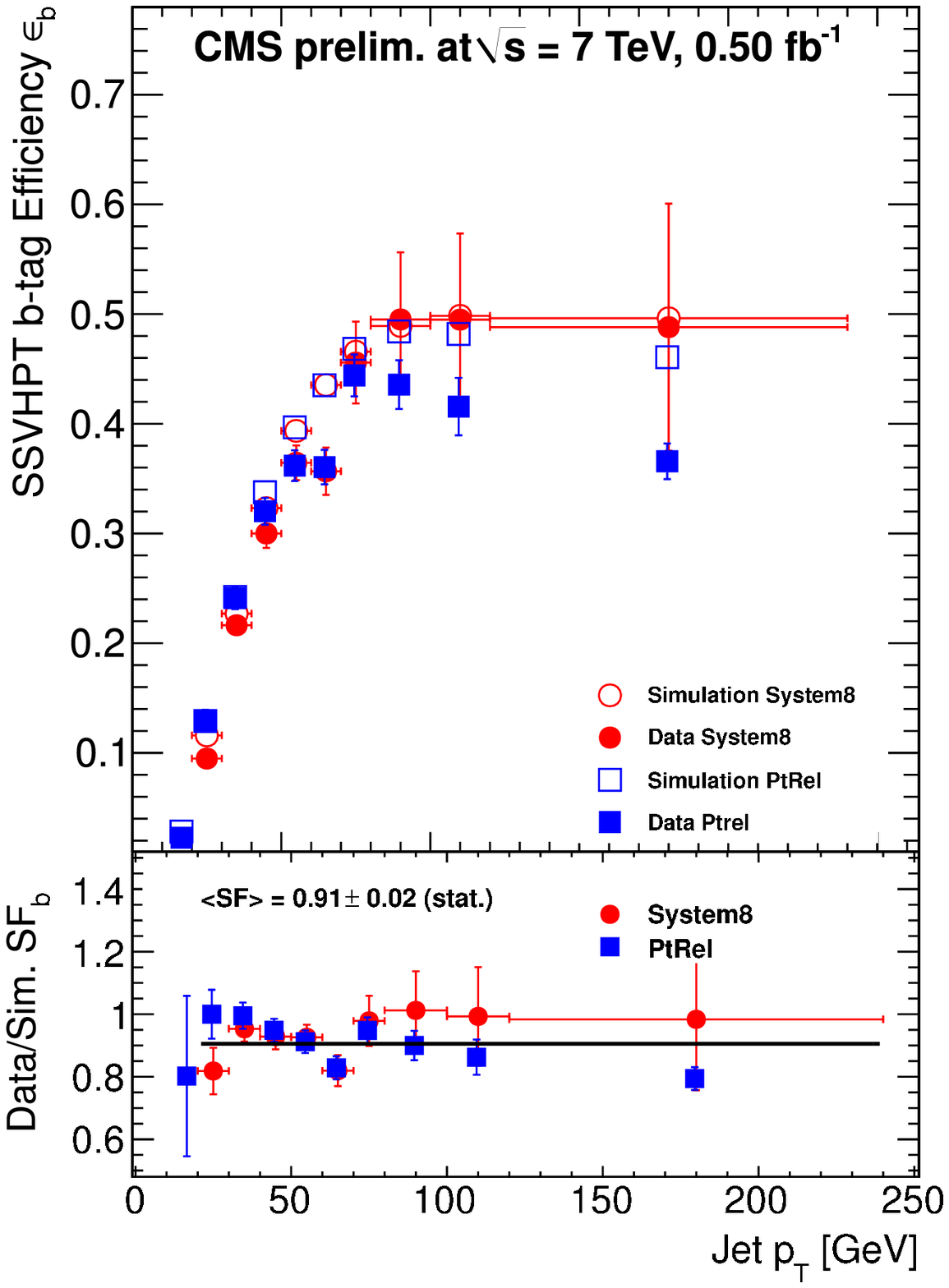}
\caption{b-tagging for the TCHEL (left panel)  and SSVHPT (right panel) taggers as a function of muon-jet $p_{T}$. Both lower panels show data/MC scale factors.}
\label{eff_plot}
\end{figure}

This section contains the measured b-tagging efficiencies, with the use of the $p_{Trel}$ and the System8 method, parametrized in jet $p_{T}$. Table~\ref{measured_eff} contains the efficiency values along with the statistical and systematic uncertainty. The sources of systematic uncertainties are described in the next section. The left panel of Fig.~\ref{eff_plot} shows that there is good agreement between the two methods and also with Monte Carlo (MC) generator level information. However, the plot on the right panel shows considerable disagreement in the high $p_{T}$ region. This can be attributed to low statistics in high $p_{T}$ bins when a high purity tight operating point is used. 

In all cases, the ratio of data to MC generator level information (scale factor, SF) is calculated. The scale factor is a measure of the departure from ideality, hence they are expected to be close to $\sim$ 1. The scale factors along with the efficiencies are used for various physics analysis involving b-jets. In Table~\ref{sf_eta} the scale factors are parametrized as a function of the pseudorapidity, $\eta$. No major variation with respect to $\eta$ is observed. 

\begin{table}[h]
\caption{Measured b-tagging efficiencies and data/MC scale factors for several b-tagging algorithms. Uncertainties are statistical for $\epsilon_{b}^{tag}$ and statistical+systematic for $SF_{b}$.}
\begin{center}
\begin{tabular}{ccccc}
b-tagger & $\epsilon_{b}^{tag}$ & $SF_{b}^{tag}$ & $\epsilon_{b}^{tag}$ &  $SF_{b}^{tag}$\\
50-80 GeV &  PtRel &   Ptrel &           System8     &                System8\\
\hline
JPL & ~0.82 $\pm$ 0.01 & ~0.97 $\pm$ 0.01 $\pm$ 0.05  &  ~0.85 $\pm$ 0.02 &   ~1.00 $\pm$ 0.02 $\pm$ 0.07\\
TCHEL & ~0.76 $\pm$ 0.01 & ~0.95 $\pm$ 0.01 $\pm$ 0.05  &  ~0.77 $\pm$ 0.01 &  ~ 0.96 $\pm$ 0.02 $\pm$ 0.05\\
TCHEM & ~0.63 $\pm$ 0.01 &  ~0.93 $\pm$ 0.02 $\pm$ 0.06  &  ~0.63 $\pm$ 0.02 &   ~0.93 $\pm$ 0.02 $\pm$ 0.07\\
TCHPM & ~0.48 $\pm$ 0.01 &  ~0.92 $\pm$ 0.02 $\pm$ 0.05  &  ~0.49 $\pm$ 0.01 &   ~0.93 $\pm$ 0.03 $\pm$ 0.09\\
SSVHEM & ~0.62 $\pm$ 0.01 &  ~0.95 $\pm$ 0.02 $\pm$ 0.07  &  ~0.60 $\pm$ 0.01 &   ~0.94 $\pm$ 0.02 $\pm$ 0.06\\
SSVHPT & ~0.38 $\pm$ 0.01 &  ~0.89 $\pm$ 0.02 $\pm$ 0.06  &  ~0.37 $\pm$ 0.01 &   ~0.90 $\pm$ 0.03 $\pm$ 0.05\\
TCHPT & ~0.36 $\pm$ 0.01 &  ~0.88 $\pm$ 0.02 $\pm$ 0.05  &  ~0.37 $\pm$ 0.01 &   ~0.88 $\pm$ 0.03 $\pm$ 0.07\\
\hline
\end{tabular}
\end{center}
\label{measured_eff}
\end{table}

\begin{table}[h]
\caption{Measured data/MC scale factors for several b-tagging algorithms in the overall jet $p_{T}$ range from 20 to 240 GeV for pseudorapidity $|\eta| <$ 2.4, $|\eta| <$ 1.2, 1.2 $< |\eta| <$ 2.4. Uncertainties are statistical for $\epsilon_{b}^{tag}$ and statistical+systematic for $SF_{b}$. Both $p_{Tel}$ and System8 provide values compatible with each other.}
\begin{center}
\begin{tabular}{ccccc}
b-tagger  & $SF^{tag}_{b}$ & $SF^{tag}_{b}$ & $SF^{tag}_{b}$\\
20-240 GeV &    $|\eta| <$ 2.4 &       $|\eta| <$ 1.2  &  1.2 $< |\eta| <$ 2.4 \\
\hline
JPL & ~0.99 $\pm$ 0.01$\pm$ 0.10 & ~~0.99 $\pm$ 0.01 $\pm$ 0.10  & ~~0.98 $\pm$ 0.01$\pm$ 0.10\\
TCHEL & ~0.95 $\pm$ 0.01$\pm$ 0.10 & ~~0.95 $\pm$ 0.01 $\pm$ 0.10  & ~~0.95 $\pm$ 0.02$\pm$ 0.10\\
TCHEM & ~0.94 $\pm$ 0.01$\pm$ 0.09 & ~~0.94 $\pm$ 0.01 $\pm$ 0.09  & ~~0.93 $\pm$ 0.02$\pm$ 0.09\\
TCHPM & ~0.91 $\pm$ 0.01$\pm$ 0.09 & ~~0.91 $\pm$ 0.02 $\pm$ 0.09  & ~~0.90 $\pm$ 0.03$\pm$ 0.09\\
SSVHEM & ~0.95 $\pm$ 0.01$\pm$ 0.10 & ~~0.95 $\pm$ 0.01 $\pm$ 0.10  & ~~0.93 $\pm$ 0.02$\pm$ 0.09\\
SSVHPT& ~0.90 $\pm$ 0.02$\pm$ 0.09 & ~~0.89 $\pm$ 0.02 $\pm$ 0.09  & ~~0.90 $\pm$ 0.03$\pm$ 0.09\\
TCHPT & ~0.88 $\pm$ 0.02$\pm$ 0.09 & ~~0.88 $\pm$ 0.02 $\pm$ 0.09  & ~~0.87 $\pm$ 0.03$\pm$ 0.09\\
\hline
\end{tabular}
\label{sf_eta}
\end{center}
\end{table}

\newpage

\subsection{Systematic Uncertainties}

Several sources of systematic uncertainties were identified. Some of these were method dependent, while most of the systematic uncertainties are common to both methods. A $p_{Trel}$-method specific systematic uncertainty was from the mismodeling of the light jet $p_{Trel}$ spectra. This was determined by constructing a collision data sample with the application of basic kinematic cuts and quoting the disagreement between data and simulations as the uncertainty. For the System8 method, the dependence on various event topologies, was a source of uncertainty. Essentially, this allowed one to vary the MC parameters in the equations and obtain the uncertainty due to their variation. Also, the $p_{Trel}$ cut was changed from 0.5 to 1.2 GeV to estimate the uncertainty due to this requirement. The rest of the sources of systematic uncertainty discussed below are applicable to both methods. The average systematic uncertainty varied between 6$\%$-7$\%$. The contributions from each source of systematic uncertainty is listed in Table~\ref{sys_ptrel} for the $p_{Trel}$ method and in Table~\ref{sys_s8} for the System8 method.

\begin{itemize}

\item Pile-up: The distribution of primary vertices from simulations were reweighted to match data. Systematic uncertainties were estimated by constructing two samples with high and low pileup regions. 

\item Away jet tagger: Dependency of the away-jet tagger on btagging efficiency was obtained by changing the taggers and the operating points.

\item Muon $p_{T}$: Muon $p_{T}$ cut was varied from its central value at 5 GeV to 7 and 10 GeV.

\item Gluon splitting: To account for the error in mismodeling gluon to $b \bar{b}$ pairs. The number of events with gluon splitting was artificially changed by a factor of two to calculate this effect.

\item Closure test: The methods were checked for self-consistency. The difference between the efficiency measurement from data and simulation was quoted as the uncertainty.

\end{itemize}

\begin{table}[h]
\caption{Sources of systematic uncertainties for the Ptrel method.}
\begin{center}
\begin{tabular}{cccccc}
b-tagger  & pile-up &  away jet & muon p$_{T}$& light & $g\rightarrow b{\bar b}$\\
\hline
JPL & ~0.2$\%$ & ~3.0$\%$ & ~2.3$\%$ & ~2.8$\%$ & ~0.3$\%$\\
TCHEM & ~2.4$\%$ & ~3.6$\%$ & ~1.5$\%$ & ~3.3$\%$ & ~0.2$\%$\\
TCHEM & ~0.9$\%$ & ~5.1$\%$ & ~1.5$\%$ & ~3.7$\%$ & ~0.1$\%$\\
TCHPM & ~1.8$\%$ & ~3.3$\%$ & ~2.6$\%$ & ~3.4$\%$ & ~0.4$\%$\\
SSVHEM & ~1.4$\%$ & ~5.8$\%$ & ~1.9$\%$ & ~3.4$\%$ & ~0.6$\%$\\
SSVHPT & ~1.1$\%$ & ~4.8$\%$ & ~2.8$\%$ & ~3.4$\%$ & ~0.6$\%$\\
TCHPT & ~0.6$\%$ & ~4.3$\%$ & ~2.3$\%$ & ~3.7$\%$ & ~0.3$\%$\\
\hline
\end{tabular}
\end{center}
\label{sys_ptrel}
\end{table}

\begin{table}[h]
\caption{Sources of systematic uncertainties for the System8 method}
\begin{center}
\begin{tabular}{ccccccc}
b-tagger  & pile-up &  away jet & muon p$_{T}$ & p$_{Trel}$ & $g\rightarrow b{\bar b}$ & sample\\
\hline
JPL & 5.1$\%$ & 1.3$\%$  & 0.8$\%$  &  2.2$\%$  &  0.1$\%$  &  3.8$\%$\\
TCHEM & 3.3$\%$  &  2.4$\%$ & 2.8$\%$  &  0.9$\%$  &  0.6$\%$  & 1.9$\%$\\
TCHEM & 5.8$\%$ & 2.6$\%$ & 0.9$\%$  &  2.0$\%$  &  0.7$\%$  &  2.4$\%$\\
TCHPM & 4.8$\%$ & 3.9$\%$ & 4.9$\%$  & 1.7$\%$  &  2.1$\%$  &  4.0$\%$\\
SSVHEM & 3.5$\%$ & 4.6$\%$ & 0.4$\%$  & 1.8$\%$  &  0.2$\%$  &  3.0$\%$\\
SSVHPT & 1.2$\%$ & 2.9$\%$ & 2.8$\%$  &  2.4$\%$  &  0.2$\%$  &  3.0$\%$\\
TCHPT & 3.5$\%$ & 3.1$\%$ & 4.0$\%$  &  2.8$\%$  &  2.5$\%$  &  2.5$\%$\\
\hline
\end{tabular}
\end{center}
\label{sys_s8}
\end{table}

\section{Cross-checks with $t{\bar{t}}$ events}
In the standard model, $t$ decays to $Wb$ at least 99.8$\%$ of the time.  The measurement of heavy flavor content, can lead to a measurement of $R_{b}  = \big(\frac{B(t \rightarrow Wb )}{B(t \rightarrow Wq)}\big)$, where $q$ is any down type quark. $R_{b}$, if assumed to be 1, can be used to extract the b tagging efficiency. Several methods were used for the determination of b-tagging efficiencies:
\begin{itemize}
\item The Profile Likelihood Ratio method : This method uses dilepton $t{\bar{t}}$ events. The distribution of jet multiplicity versus b-tagged jet multiplicity in dilepton $t{\bar{t}}$ events is used to construct a likelihood function.  
\item The $R_{b} $ method : The methods also replies dilepton $t{\bar{t}}$ events. The observed b-tagged jets is proportional to the fraction of b-jets present, the proportionality factor being $\epsilon_{b}^{tag}$. The number of b-tagged jets is modeled probabilistically using $\epsilon_{b}^{tag}$ and $\epsilon^{mistag}$ for dilepton $t\bar{t}$ events.
\item The Flavor Tag Consistency Method :  lepton+jets $t{\bar{t}}$ events from top decays are used as input to this method. The procedure requires consistency between observed and expected number of identified jets in an event in $t\bar{t}$ lepton+jets decays . A dedicated likelihood function is built based on $\epsilon_{b}^{tag}$, $\epsilon_{c}^{tag}$ and $\epsilon_{mistag}$, $t{\bar{t}}$ cross section and acceptance obtained from simulations.  
\item The Simultaneous Heavy Flavor and Top method : This method also uses lepton+jets $t{\bar{t}}$ events. $\epsilon_{b}^{tag}$ is obtained from two-dimensional fit with the number of jets and the invariant mass of the tracks forming the secondary vertex.  
\end{itemize}
All of these methods give efficiency values compatible with Ptrel and System8 methods and are also consistent with each other.

\section{Estimation of mis-tag rate with Negative Taggers}

\begin{figure} 
 \includegraphics[height=3.0in, width=4.0in]{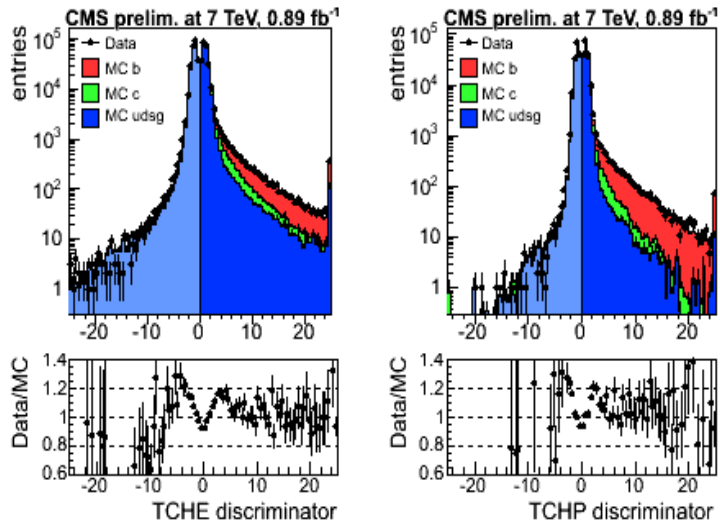}
 \caption{Signed $b$-tag discriminators in data (dots) and simulation of light flavor jets (blue), c-jets(green) and b-jets (red area) with a $p_{T}$ threshold of 30 GeV. }
 \label{neg_tag} 
\end{figure}

 The mis-tag rate is obtained from tracks with negative impact parameters or secondary vertices with negative decay lengths. The TC discriminators are plotted in Fig.~\ref{neg_tag}. The negative IPs are ordered from the most negative upwards. The ordering on the positive side remains unchanged. The negative taggers are used in the same way as the current b-tagging algorithms. The mis-tag rate is evaluated as: ${\varepsilon}_{data}^{mistag} = {\varepsilon}^{-}_{data}.R^{light} $, where $\varepsilon^{-}_{data}$ is the negative tag rate in data and $R_{light} = \varepsilon_{MC}^{mistag}/\varepsilon_{MC}^{-}$ is the ratio between the light flavor mis-tag rate and negative tag rate of all jets in the simulation. The measured mis-tag rates are in Table~\ref{mistag}. The light jet scale factors are also included.  
 
\subsection{Systematic Uncertainties}

The following sources of systematic errors were taken into consideration:\\
\begin{itemize}

\item b and c fractions: The b+c flavor fraction is varied in the QCD simulations and a systematic uncertainty is obtained on $R_{light}$ (1.9$\%$).
\item Gluon fraction: Uncertainty is extracted from comparison of simulation with data (0.2$\%$).
\item Long lived $K_{s}^{0}$ and $\Lambda$ decays (displaced vertices) and photon conversion and nuclear interactions (2.0$\%$). QCD simulation events are re-weighted to take into account the observed yields of $K_{s}^{0}$ and $\Lambda$ in data since these processes involve displaced vertices.
\item Mismeasured tracks: Spurious tracks increase the number of positive over negative tags (0.3$\%$). 
\item Sign flip: The ratio of the number of negative and positive tagged jets is computed in a muon-jet sample with a larger than 80$\%$ b purity (4.3$\%$).
\item Event sample (dominant systematic): Using jets originating from different event topologies. Dominant systematic (10$\%$).
\item Pile up: Uncertainty estimated in the same way as described above (0.7$\%$). 
\end{itemize}

\begin{table}[h]
\caption{Mis-tag rate and data/MC scale factor for different b-taggers with $p_{T}$ between 50 and 80 GeV. The statistical+systematic uncertainties are quoted.}
\begin{center}
\begin{tabular}{ccccc}
b-tagger  & ~~~~~mis-tag rate (${\varepsilon}_{data}^{mistag}$)~~~~~ & Scale Factor for light jets (${\varepsilon}_{data}^{mistag}/{\varepsilon}_{MC}^{mistag}$)\\
\hline
JPL & ~~~0.077 $\pm$ 0.001$\pm$ 0.016 & 0.98 $\pm$ 0.01 $\pm$ 0.11\\  
TCHEL & ~~~0.128 $\pm$ 0.001$\pm$ 0.026 & 1.11 $\pm$ 0.01 $\pm$ 0.12\\
TCHEM & ~~~0.0175 $\pm$ 0.0003$\pm$ 0.0038 & 1.21 $\pm$ 0.02 $\pm$ 0.17\\
SSVHEM & ~~~0.0144 $\pm$ 0.0003$\pm$ 0.0029 & 0.91 $\pm$ 0.02 $\pm$ 0.15\\
SSVHPT & ~~~0.0012 $\pm$ 0.0001$\pm$ 0.0002 & 0.93 $\pm$ 0.09 $\pm$ 0.12\\
TCHPT & ~~~0.0017 $\pm$ 0.0001$\pm$ 0.0004 & 1.21 $\pm$ 0.10 $\pm$ 0.18\\
\hline
\end{tabular}
\label{mistag}
\end{center}
\end{table}

\section{Conclusion}

Several methods have been used to obtain the tagging efficiency of b jets using an integrated luminosity of 0.50 to 0.89 fb$^{-1}$ collected by the CMS experiment in 2011. The data/MC scale factor is measured with an uncertainty of 10$\%$ for b jets with $p_{T}$ up to 240 GeV. For light flavor jets with $p_{T}$ up to 500 GeV the mis-tag rate is measured with an uncertainty of 10-20$\%$. B-tagging efficiencies are cross checked with independent analyses using $t{\bar{t}}$ events. B tagging is of crucial importance in events with topologies involving b quarks \cite{single_top}.

\end{document}